# ACTreS: Analog Clock Tree Synthesis

Bilgiday Yuce, H. Fatih Ugurdag, Iskender Agi,
Gokhan Guner, Vahap Baris Esen, Seyrani Korkmaz,
I. Faik Baskaya, Günhan Dündar

*Abstract*—This paper describes a graph-theoretic formalism and a flow that, to a great extent, automate the design of clock trees in Sampled-Data Analog Circuits (SDACs). The current practice for clock tree design of SDACs is a manual process, which is time-consuming and error-prone. Clock tree design in digital domain, however, is fully automated and is carried out by Clock Tree Synthesis (CTS) software. In spite of critical differences, SDAC clock tree design problem has fundamental similarities with its digital counterpart. We exploited these similarities and built a design flow and tool set, which uses commercial digital CTS software as an intermediate step. We will explain our flow using a 0.18 micron 10-bit 60 MHz 2-stage pipelined differential-input flash analog-to-digital converter as a test circuit.

## I. INTRODUCTION

SAMPLED-Data Analog Circuits (SDACs) are used in many applications. Examples of SDACs are switched-capacitor filters, instrumentation amplifiers, voltage-to-frequency converters, balanced modulators, peak detectors, oscillators, ADCs, DACs, and programmable capacitor arrays. SDACs require clock signals to operate, just like digital circuits. However, unlike digital circuits, they require many phases of the same clock, and the relationships between the phases are critical. There are 3 possible relationships between clock phases, namely, non-overlapping, specific order, and enclosing (see Fig. 1).

Currently, design of clock phase generation and distribution circuitry for SDACs is a lengthy and ad hoc manual process, and hence, can be source of errors as was the case with an industry design team, which the last author belonged to. The company had to do a costly design respin of an SDAC chip due to a design mistake in the clock circuitry.

This work was funded by The Scientific & Technological Research Council of Turkey (TÜBITAK) under grant no. 110E061. During this work, B. Yuce was both with Bogazici University, Bebek 34342, Istanbul, Turkey and Ozyegin University, Cekmekoy 34794, Istanbul, Turkey. H. F. Ugurdag is with Ozyegin University (email: fatih.ugurdag@ozyegin.edu.tr). During this work, Dr. Iskender Agi was with Agi Consulting, Redwood City, CA 94062, USA. G. Guner was with Ozyegin University. V. B. Esen and S. Korkmaz were with Bogazici University. Prof. I. F. Baskaya and Prof. G. Dündar are with Bogazici University.

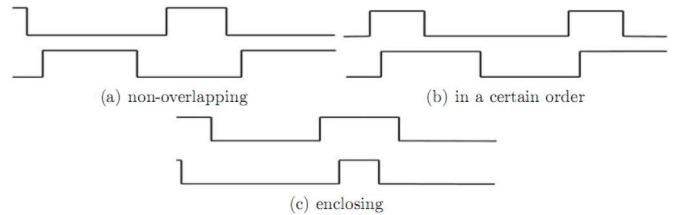

Fig. 1. The three types of clock relationships.

The design of clock circuitry for a given source clock in digital chips (a.k.a. clock trees) is a fully automated process realized by Clock Tree Synthesis (CTS) software. This led us to the rightful question of "Can we automate the design of clock circuitry for SDACs as well?" We have automated most of the process. However, more importantly, we were able to come up with a systematic design flow, which can be adhered to even when the design is done manually or partly manually (i.e., semi-automation), so that the process is not ad hoc and the design can be checked against certain quality metrics.

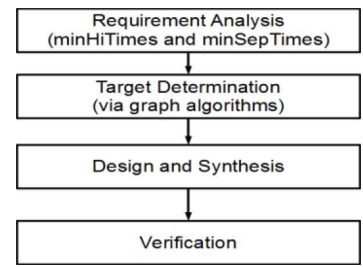

Fig. 2. Steps of ACTreS.

We assume that the design of the SDAC has either fully or partly finished before the design of clock circuitry (which is a clock tree in a loose sense and will be called so from this point on) part of it. Our design flow starts by interviewing the analog designer and capturing the clock *requirements* (i.e., phase relationships) in the form of a directed graph (*digraph*) where the nodes are clock edges and arrows are time offsets between edges. This digraph can be thought of as a set of *constraints* and defines a region in the solution space. The next step is picking a *target* point in that solution space, which involves solving a *scheduling* problem. Then comes the *design* process followed by a *verification* step that is supposed to *measure the quality* of the design with our clock trees versus some reference. Our design process leverages a unique two-level clock tree *architecture* (an "intrinsic clock tree" generates the phases, each of which is distributed to end points, i.e., switches, through a separate "extrinsic clock tree"). The design process also involves generating the appropriate scripts and parameter files and then running Cadence SoC Encounter (with its digital CTS capability) to produce a placed and routed overall circuit, within which the SDAC (i.e., analog macro) is a blackbox. To summarize, the critical stages of our flow is shown in Fig. 2.

Speaking of verification, we also have several small verification tools that we use at intermediate points. That is, we are able to automatically check requirements files (i.e., constraints) and targets for sanity – whether they have any internal inconsistencies.

To build a solution like this and eventually test it, one needs an SDAC design problem. To that end, we designed a 0.18 micron 10-bit 60 MHz 2-stage pipelined differential-input flash Analog-to-Digital Converter (ADC) [1]. We check whether our Analog Clock Tree Synthesis (ACTreS) is successful or not by comparing the Effective Number Of Bits (ENOB) of the design with synthesized clock trees versus the ENOB of the design using targeted clock phases at the three characterization corners (i.e., slow, typical, fast). We have completed the layout of the ADC then obtained ENOB values through Spice simulation and post-processing through MATLAB.

The next section reviews the literature. There is a separate section for every stage of ACTreS in Fig. 2 between Sections III and VI. Section VII presents the results based on our test circuit and concludes the paper.

## II. PREVIOUS WORK

There is a plethora of work on digital CTS, a recent survey of which is given in [2]. However, there is no work specifically on SDAC clock tree synthesis. There are works such as [3] and [4] that attempt to automate the complete circuit (even layout) design process for a particular subclass of SDACs. However, they do not include much detail in terms of CTS and how it can be generalized. There are also works that focus on the generation of two non-overlapping clocks ([5], [6]). Two non-overlapping clocks is enough for some SDACs but for many SDACs we need more phases such as multi-rate SDACs [7], some SDCAs for power electronics applications [8], and SDACs with interleaved operation [9], which is a subclass that our test circuit also falls in.

The normal digital CTS paradigm is about distributing a clock to many end points with zero skew if possible. That is useful in our problem in "distributing" the generated clock phases but does not help generate the various clock phases an SDAC needs. Thanks to the idea of "clock-skew scheduling" [10], digital CTS tools have the capability of generating deliberate timing offsets between end points and hence can be used to generate different phases of a main clock.

Note that a preliminary and condensed version of our work was published in [11]. At the time we were trying to clock our test circuit at 180 MHz. However, we had some analog design challenges (not in the CTS part) that we could not overcome. To successfully complete the analog design of our test circuit, we had to lower the clock frequency to 60 MHz.

## III. REQUIREMENT ANALYSIS

The first step of our clock tree design flow is requirement analysis, in which we examine the SDAC circuit to figure out its clock requirements/constraints and express them in a text file. The requirement analysis we do for an SDAC can be broken up into four steps (Fig. 3), each of which is devoted a subsection in this section.

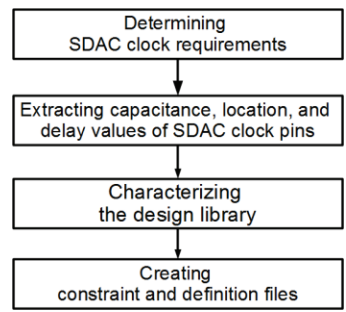

Fig. 3. Steps of Requirement analysis.

As it is stated in Section I, our assumption is that the SDAC design is fully or partially completed before starting requirement analysis. Hence, the requirement analysis requires interaction with the designer of the SDAC to understand the clock requirements. Therefore, it is useful to explain our test circuit before proceeding further.

In this paper, we use a *10-bit 60 MHz 2-stage pipelined differential-input flash analog-to-digital converter* (*ADC*) as the test circuit. The architecture of our test circuit is provided in Fig. 4.

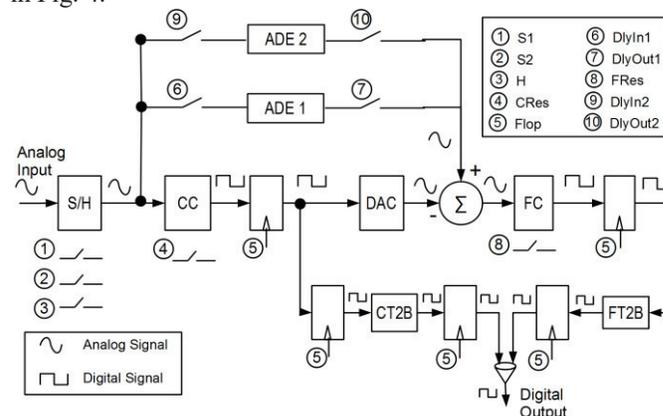

Fig. 4. Block diagram of the test circuit.

Our test circuit consists of functional blocks and switches that control them. Note that each switch symbol in Fig. 4 represents a group of switches driven by a particular clock phase. A conversion of our ADC consists of 12 operations:

1. *Smp*: The input signal is sampled by *Sample and Hold* (*S/H*) block. This operation is controlled by switches $S_1$ (switch 1) and $S_2$ (switch 2).
2. *H*: The sample is sent to the circuit for processing. This operation is done when switch 3 is closed.
3. *CC*: The sample is compared with reference values by *Coarse Comparator* (*CC*) block. This operation determines the most significant 5 bits of the ADC's output. It is carried out when switch 4 is open.
4. <u>CRes</u>: The output of coarse comparator block is reset to zero via switch 4 (when it is closed).
5. <u>$DlyIn_1/DlyIn_2$</u>: is the first operation, which is related to *Analog Delay Element$_1$* ($ADE_1$) /$ADE_2$. The value sampled by S/H is transferred to the output of the corresponding ADE block at the end of this operation. Then, the transferred value is used by *Subtractor* (*SUB*) and *Digital to Analog Converter* (*DAC*) blocks. We use ADE blocks in a ping pong buffer scheme. $ADE_1/ADE_2$ receives the next sample value while the current sample value, which resides at the output of $ADE_2/ADE_1$ is processed by SUB and DAC blocks. Due to ping pong buffering, the CC stage (1$^{st}$ stage of the ADC) and the FC stage (2$^{nd}$ stage) can operate in parallel on different samples. Switches 6 and 9 control this operation.
6. <u>$DlyOut_1/DlyOut_2$</u>: is the other operation related to ADE. During this operation, the output of the corresponding ADE block holds its value, which is then processed by SUB and DAC. It is controlled by switches 7 and 10.
7-8. <u>DAC+SUB</u>: Analog input signal for the fine comparison is obtained by DAC and SUB blocks. The S/H output is first converted into an analog signal using the DAC. Then, this converted signal is subtracted from the output of ADE using the SUB. The result of the subtraction is the input signal of the fine comparison stage. This operation does not require a clock phase.
9-10. <u>FRES+FC</u>: are similar to their coarse comparison counterparts. However, they use *fine comparator* (*FC*) block and determine the least significant 5 bits of our ADC's output. And they are controlled with switch 8.
11-12. <u>CT2B+FT2B</u>: 31-bit thermometer codes produced by CC and FC blocks are converted to 5-bit binary codes by CT2B and FT2B blocks, respectively. As result, we obtain a 10-bit digital ADC output.

TABLE I
HIGH LEVEL DEFINITION OF CLOCK PHASES

| Switch Number | Phase Name | Duty Cycle | Period |
|---|---|---|---|
| - | clk | $T_{clk}/2$ | $T_{clk}$ |
| 1 | $S_1$ | $T_{clk}/2$ | $T_{clk}$ |
| 2 | $S_2$ | $T_{clk}/2$ | $T_{clk}$ |
| 3 | H | $T_{clk}/2$ | $T_{clk}$ |
| 4 | CRes | $T_{clk}/2$ | $T_{clk}$ |
| 5 | Flop | $T_{clk}/2$ | $T_{clk}$ |
| 6 | DlyIn1 | $T_{clk}/2$ | $2*T_{clk}$ |
| 7 | DlyOut1 | $T_{clk}$ | $2*T_{clk}$ |
| 8 | FRes | $T_{clk}/2$ | $T_{clk}$ |
| 9 | DlyIn2 | $T_{clk}/2$ | $2*T_{clk}$ |
| 10 | DlyOut2 | $T_{clk}$ | $2*T_{clk}$ |

A. *Determining SDAC Clock Requirements*

Timing diagram of our test circuit is given in Fig. 5, which is obtained by interviewing the analog designer. We divide each clock cycle into 2 time-slots in Fig. 5. Every row of Fig. 5 corresponds to a time-slot and it contains ADC operations, which are carried out in this time-slot. We include the samples *n*, *n+1*, and *n+2* in Fig. 5. As it can be seen, throughput of our ADC is 2 time-slots while its latency is 6 time-slots. Moreover, our ADC processes three input samples at the same time as it is shown in time-slots 5 and 6 of Fig. 5.

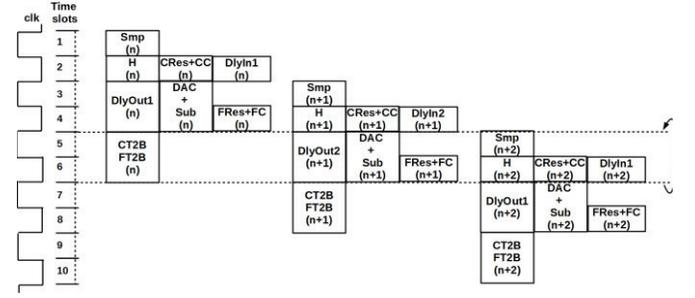

Fig. 5. Timing diagram for the test circuit.

Figure 5 also intrinsically contains high-level definitions (i.e., period and duty cycle values) of the required clock phases. For instance, clock phases that control the *Smp* operation should be high when the main clock (*clk*) is low. As it can be seen from Fig. 5, period values of these clock phases should be equal to period of the clk ($T_{clk}$), and their duty cycles should be close to $T_{clk}/2$. Similarly, we obtain the list of all high-level period and duty cycle values, which is given in Table I.

After interviewing the analog designer and considering the timing diagram, the following clock requirements (i.e., requirements list) were determined: (Note that 'NO' represents non-overlapping, 'w/in' is within, and 'f' and 'r' are for falling and rising edges, respectively.)

- $S_2$ w/in $S_1$ (1r -> 2r and 2f -> 1f)
- $S_1$ and H NO (1f -> 3r and 3f -> 1r)
- 'negedge CRes' before 'posedge Flop' (4f -> 5r)
- 'negedge FRes' before 'posedge Flop' (8f -> 5r)
- DlyIn1 w/in H (3r -> 6r and 6f -> 3f)
- DlyIn2 w/in H (3r -> 9r and 9f -> 3f)
- 'Flop posedge' before 'H negedge' (5r -> 3f)
- 'DlyOut1 posedge' before 'DlyIn1 negedge' (7r -> 6f )
- 'DlyOut2 posedge' before 'DlyIn2 negedge' (10r ->9f)
- 'Flop posedge' before 'DlyOut1 posedge' (5r -> 7r)
- 'Flop posedge' before 'DlyOut2 posedge' (5r -> 10r)
- 'CRes posedge' before 'H negedge' (4r -> 3f)
- 'FRes posedge' before 'H negedge' (8r -> 3f)

As it can be seen from the requirements list, a clock requirement can specify one of:
- the minimum time between the rising and falling edges of the same clock phase (i.e., *minimum high time - minHiTime*),
- the minimum time between edges of two different clock phases (i.e., *minimum separation time - minSepTime*), or
- the minimum time during which two different clock phases should be high (i.e., *minimum duration time - minDurTime*).

B. *Extraction*

Next step is extracting capacitance, location, and delay values for the clock pins of the analog circuit using the analog design environment. These values will be used in *Design and Synthesis* step of ACTreS.

*C. Characterization*

Further down our flow, we will need to know how fast our standard-cell library is at slow, typical, and fast PVT (Process, Voltage, Temperature) relative to each other. Although these values can vary depending on the circuit, an approximate fudge factor can be computed and used independent of the circuit in consideration. For this purpose, we connected each buffer and inverter cell in a binary tree configuration, and we measured the delay of each binary tree, via static timing analysis, for different PVT corners. Considering the worst delay values that we obtained, we can claim that:

- The standard-cell library is 1.6 times faster at the fast corner compared to the normal (i.e., typical) corner ($k_{fn}$ = 1.6).
- The standard-cell library is 2.5 times faster at the fast PVT corner compared to the slow corner ($k_{fs}$ = 2.5).

*D. Creating Constraint and Definition Files*

The final step of the requirement analysis is producing constraint and definition files for the later stages of our flow.

*Constraint file* is created by specifying a minimum time, which has to be satisfied by the analog circuit in any case, for each requirement in the requirements list. This process can be thought of as converting the high-level phase relations in Fig. 5 into more stringent relations. In addition to the requirements of the requirement list, constraint file also includes *minHiTime* and *minDurTime* values for the *critical clock phases*. We define a critical clock phase as a clock phase, of which minHiTime/minDurTime value is essential for the performance of the circuit. We utilize the mathematical error modeling of the analog circuit and the analog designer's experience to determine the critical clock phases as well as their minHiTime/minDurTime values. For our design, S2, DlyIn$_1$/DlyIn$_2$, DlyOut$_1$/DlyOut$_2$ are the critical clock phases.

The last step of creating the constraint file is to determine the equivalent clock phases in order to create a simpler constraint file. We call two clock phases equivalent if putting one of these phases into the constraint file is enough to define the timing constraints of both phases. For instance, CRes and FRes phases have exactly the same requirements. Similarly, DlyIn1 and DlyIn2 clock phases have the same requirements. Moreover, DlyOut1 has the same requirements with the inverted version of DlyOut2. Hence, we can state that:

- CRes is equivalent to FRes
- DlyIn1 is equivalent to DlyIn2
- DlyOut1 is equivalent to DlyOut2

Taking these identical phases into account, we are going to use CRes (phase 4), DlyIn (phase 6), and DlyOut (phase 7) to represent these equivalent phases throughout this paper. As a result, we obtain the constraint file (Fig. 6a) and realistic clock phase waveforms, which are shown in Fig. 6b.

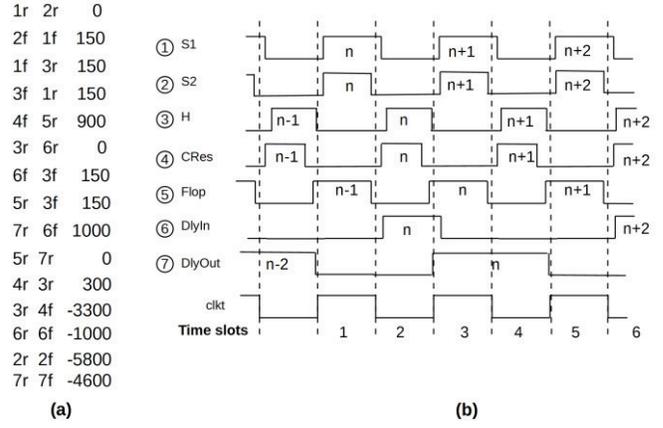

Fig. 6. (a) Constraint file. (b) Realistic clock phase waveforms.

As it is seen from Fig. 6a, the constraint file has the following format:

<1st phase no.> <edge> <2nd phase no.><edge> <t$_2$–t$_1$>

Here, t$_2$–t$_1$ denotes the timing distance between two edges.

*Definition file* is a simple input file that holds the duty cycle and period value for each clock phase. It has the following format:

<phase no.> <phase name> <duty cycle> <period> <inv>

In the definition file (Fig. 7a), duty cycle and period values are represented as multiples of $T_{clk}/2$. The 'inv' column in the format determines whether a phase (0) or its inverted version (1) is produced.

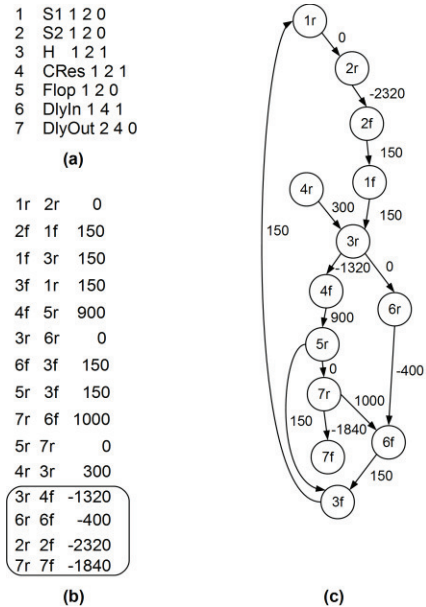

Fig. 7. (a) Definition file. (b) Updated constraint file. (c) Graph of (b).

*E. Updating the Constraint File and Observing its Graph*

The analog circuit should satisfy the constraints for all PVT corners after synthesis of the clock tree. Different constraint types (minSepTime, minHiTime, minDurTime) in the constraint file take their lowest values at different PVT corners. For example, minSepTime constraints get their smallest values at the fast PVT corner. If the value of a minHiTime or minDurTime constraint is smaller than $T_{clk}/2$ its value is smallest at fast PVT corner, otherwise, its value is smallest at slow PVT corner. Hence, we should choose a PVT corner for synthesis and we should project the constraints to the chosen PVT corner.

It is clear that the design will not work properly if some minSepTime constraints cannot be achieved. In the event of not satisfying some minHiTime constraints, the design will work with a lower performance. Based on this information, it is evident that satisfying the minSepTime constraints is more critical. Therefore, we need to aim for fast corner.

To update the constraint file, we divide negative minHiTime and minDurTime values of the constraint file by $k_{fs}$, which is computed in the extraction step. The resulting updated constraint file is given in Fig. 7b. In this figure, we mark the updated parts of the constraint file with a rectangle. The process of obtaining updated constraint file from the original one is handled by a Perl script.

Moreover, the constraint file can also be represented as a digraph. This enables the use of graph algorithms on the constraint file. Digraph for the updated constraint file is given in Fig. 7c. Each clock edge in the constraint file is implemented as a node, and each constraint in the file is an edge between nodes.

## IV. TARGET DETERMINATION

The updated constraint file constructed in the previous section defines a safe zone, in which we can select our clock phases. However, feeding these constraints directly to the CTS tool may result in two potential problems. First, some of the synthesized clock phases may be beyond bounds of the safe zone. Second, the performance of the overall circuit may be suboptimal. In this section, we will explain how we obtain *actual targets* that are fed to the CTS tool from our updated constraint file.

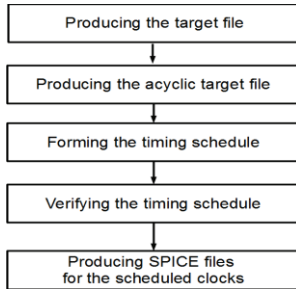

Fig. 8. Steps of Target Determination.

The actual targets guarantee that all synthesized clock phases are within the safe region. In addition, we increase the performance by selecting slack values that maximizes the minHiTime constraints.

The flow of target determination is given in Fig. 8. At the end of this flow, we obtain a *schedule file* that contains the phase differences between the main clock and clock phases in the constraint file.

In order to overcome these potential problems, we should add a *slack* value on top of each edge weight of the updated constraint file's digraph. In addition, we select slack values that maximize minHiTime constraints to increase the performance. These new constraints are stored in *target file*.

### A. Producing the Target File

The first step of target determination is creating the target file by adding slack values on top of the edge values in the digraph of the updated constraint file. In order to select the slack values, we use a graph algorithm, which tries to maximize minHiTime constraints for better performance.

This process is automatically carried out by a Perl script. The first step of creating the target file is finding the cycles in the digraph of the updated constraint file. Our Perl script uses a Python script, which is obtained from [12]. The author of [12] states that the Python script was created based on [13]. The Python script finds the cycles shown in Fig. 9.

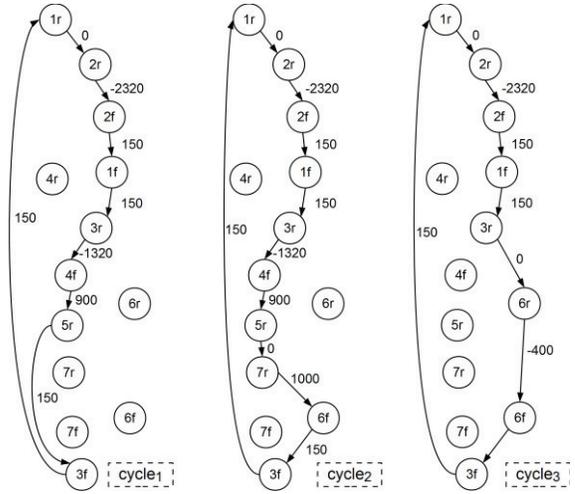

Fig. 9. Cycles of the updated constraint file.

After identifying the cycles, the target file is obtained by adding slack to the edge weights. We use a graph algorithm, whose pseudocode is given in Fig. 10a, to determine the slacks to be added. Our algorithm makes an effort to maximize minHiTime values. The algorithm works as follows:

1. The length of each cycle in the updated constraint file is calculated. The length of a cycle is the sum of the edge weights within the cycle:

$$length_1 = -2140$$
$$length_2 = -1140$$
$$length_3 = -2120$$

2. For each cycle *j*, an average slack is computed using (1):

$$avgSlack_j = length_j / \# \ nodes \ in \ j \quad (1)$$

Using (1), we obtain the following values for Fig. 9:

$$avgSlack_1 = -267.5$$
$$avgSlack_2 = -114.0$$
$$avgSlack_3 = -265.0$$

3. In this step, slack value of each minSepTime edge that is included by at least one cycle is computed. The slack value of an edge e is the maximum of the average slack values of the cycles, which contains the edge e, and the default slack value (-100 for this paper). As the default slack is greater than all of the average slack in step 2, for our circuit, slack value for an edge that is in a cycle will be -100.
4. In this step, the slack values computed in step 3 are subtracted from the corresponding cycle lengths. As a result, remaining slack values for each cycle is computed. For our circuit, we obtain the following results:

$$remainingSlack_1 = -2140 - 7(-100) = -1440$$
$$remainingSlack_2 = -1140 - 9(-100) = -240$$
$$remainingSlack_3 = -2120 - 6(-100) = -1520$$

5. In this step, slack value of each minHiTime edge $j$ that is included by at least one cycle is calculated. First, the *remaining slack per minHiTime edge* is calculated for each cycle $c$. For this purpose, We divide $remainingSlack_c$ by the number of minHiTime edges within cycle $c$. Each remaining total slack value calculated at the previous step is divided to the number of minHiTime constraints present in that loop. Each quotient is a slack candidate for the minHiTime constraints present in the loop quotient belongs to. Then, the maximum of these values is chosen as the slack value of minHiTime edge $j$.
6. Completing the previous steps, we computed slack values for all edges, which is included at least one cycle. In this step, we assign the maximum of the average slack values computed in step 2 and the default slack value into the remaining edges. For our example, we assign "max(-267.5, -114, -265, -100) = -100" to the remaining edges.

As a result, we obtain the target file given in Fig. 11.

### B. Producing the Acyclic Target File

In the scheduling step, we need an acyclic target file. Hence, we should convert our target file into an acyclic target file by deleting the edges that result in cycles. For this purpose, we use a depth-first search (DFS) based graph algorithm, of which pseudo code is given in Fig. 10b. In this algorithm, we use two variables for each node of the target file: *visited* and *onPath*. The variable visited specifies whether a node is visited. The variable onPath specifies if the current edge causes a cycle. The steps of the algorithm can be summarized as follows:

1. For each node, visited and onPath variables are initialized to zero.
2. For each node the DFS function is called.

The DFS function, which is a recursive function, works as follows:

1. If the current node is visited, it returns. Otherwise, visited and onPath variables of the node are set to one. And the function proceeds to step 2.
2. For each child of the current node, it is checked that if going to this node causes a cycle.

If going to this child node does not result in a cycle, the corresponding edge is added to the acyclic target file and the DFS function is called from this child node. If the current node has no child, its onPath variable is assigned to 0.

As a result, we obtain the acyclic target file, which is given in Fig. 12 is obtained.

```
cSlack = -100

foreach cycle {
    foreach edge in a cycle
        totalSlack[cycle] += edgeDistance;
    avgSlack[cycle] = totalSlack[cycle]/#Edge[cycle];
}

foreach cycle {
    foreach min-sep edge {
        if (slack[edge] = "")
            slack[edge] = cSlack;
        else if (slack[edge] < avgSlack[cycle])
            slack[edge] = avgSlack[cycle];
    }
}

foreach min-sep edge {
    foreach cycle
        if (min-sep edge is at cycle)
            usedSlack[cycle] += min-sep edge;
}

foreach cycle {
    totalSlack[cycle] -= usedSlack[cycle];
    avgSlack[cycle] = totalSlack[cycle]/#min-highEdge[cycle];
}

foreach cycle {
    foreach min-high edge {
        If (slack[edge] == "")
            slack[edge] = avgSlack[cycle];
        else if (slack[edge] < avgSlack[cycle])
            slack[edge] = avgSlack[cycle];
    }
}

foreach edge {
    if (slack[edge] == "")
        slack[edge] = cSlack;
}
```
(a)

```
main() {
    for every node {
        visited(node) = 0;
        onPath(node) = 0;
    }

    for every node
        DFS(node);
}

DFS(node) {
    if(visited(node))
        return;
    visited(node) = 1;
    onPath(node) = 1;

    for every child in children(node)
    {
        if(onPath(child) == 0)
            print "node child";
            DFS(child);
    }
    onPath(node) = 0;
}
```
(b)

Fig. 10. (a) Target file generation algorithm. (b) Acyclic target file generation algorithm.

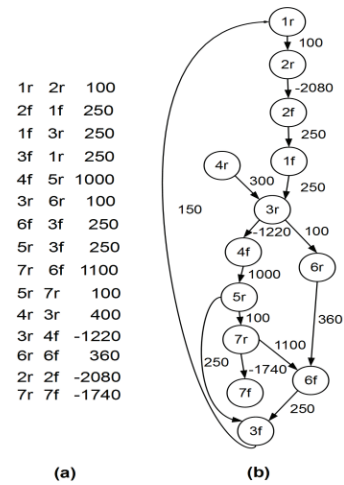

(a) (b)

Fig. 11. (a) Resulting target file. (b) Diagraph for the target file.

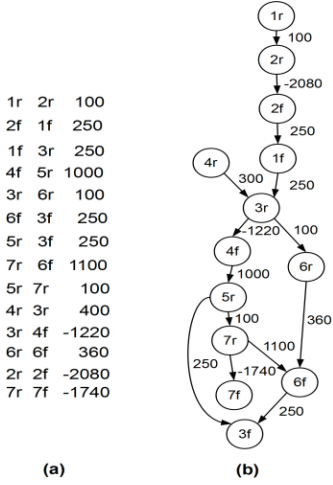

Fig. 12. (a) Resulting acyclic target file. (b) Diagraph of (a).

### C. Generating the Timing Schedule

As it will be clear in Section IV, we will generate each clock phase in the constraint file from a main clock using the Clock tree synthesis (CTS) tool. Therefore, we have to feed phase differences between the main clock and each clock edge in the constraint file to the CTS tool as input. For this purpose, we create a *timing schedule*, which contains these phase differences, from the acyclic target file. In other words, by creating the timing schedule, we specify the position of each node of the acyclic target file on the time axis. For scheduling, we use another DFS-based graph algorithm, of which pseudo code is given in Fig. 13.

The problem here is a type of "longest path" problem. The first step of the scheduling algorithm is choosing a node as the reference node and assuming its time as zero. Then, we apply the following two operations repeatedly until we schedule all nodes:

1. First, we find the root nodes, which can be reached from the scheduled nodes in both directions, using DFS: positive direction (i.e., direction pointed by arrows) and negative direction (i.e., reverse of the positive direction). This operation is called *discovery* (*D*).
2. From each discovered node in the step 1, we apply a DFS in the reverse direction of the corresponding discovery operation to reach a scheduled node. After reaching a scheduled node, search should go back to the source and schedule the nodes on the way. This operation is called *scheduling* (*S*).

In our algorithm we can apply discovery and scheduling operations concurrently. In addition, we have a root list. A node that is discovered in any step of the algorithm and the direction of corresponding DFS to be started from this root node is written into the root list. Then, the elements of the root list are read one by one and a DFS is applied in the specified direction. While applying the DFS discovery and scheduling operations are carried out. These steps are applied repeatedly until the root list becomes empty.

```
main() {
  visitedFlag = 0;
  for every node {
    visited(node) = 0;
    time(node) = UNKNOWN;
  }
  rootList = ((node0, +1), (node0, -1));
  time(node0) = 0;
  while(rootList != EMPTY) {
    visitedFlag++;
    (node, dir) = removeOneFromList(rootList);
    DFS(node, dir);
  }
}

DFS(node, dir) {
  if(visited(node) == visitedFlag)
    return;
  if((time(node) != UNKNOWN) && (node != node0) )
    return;
  visited(node) = visitedFlag;
  children = children(node, dir);
  if(children == EMPTY)
    add node to rootList with -dir;
  first = 1;
  for every child in children {
    DFS(child, dir);
    if(time(child) == UNKNOWN)
      next;
    if(first) {
      time(node) = time(child)-dir*edge(node,child);
      first = 0;
    } else {
      tmp = time(child)-dir*edge(node,child);
      if(dir*tmp < dir*time(node))
        time(node) = tmp;
    }
  }
}
```

Fig. 13. Pseudocode of the scheduling algorithm.

How the scheduling algorithm works can be explained using an example graph, which is given in Fig. 14. In our algorithm we have three kinds of nodes: *undiscovered*, *discovered*, and *scheduled*. In Fig. 14, we use dashed circles to represent undiscovered nodes, solid, gray circles to show the discovered nodes, and solid, black circles for the scheduled nodes. In addition, we use S for the scheduled nodes and D+/D- for the discovered nodes. D+ (D-) shows that the node is discovered as a result of a DFS that is applied in positive (negative) direction. The following steps are applied for the example given in Fig. 14:

1. We pick node 4 as reference. Initial root list is {4+,4-}.
2. As the first element of the root list is 4+, we apply a DFS in the positive direction from this node. And we discover the nodes 1 and 2. Our new root list is {4/+,4/-,1/-,2/-}.
3. Then, element 4/- is processed and nodes 7 and 8 are discovered. New root list is {4/+,4/-,1/-,2/-,7/+,8/+}.
4. Then, element 1/- is processed and node 1 is scheduled. In this step, we do not discover any new nodes. New root list is {4/+,4/-,1/-,2/-,7/+,8/+}.
5. Then, element 2/- is processed, node 2 is scheduled, and node 9 is discovered. New root list is {4/+,4/-,1/-,2/-,7/+, 8/+,9/+}.
6. Then, element 7/+ is processed and nodes 6 and 7 are scheduled. New root list is {4/+,4/-,1/-,2/-,7/+,8/+,9/+}.
7. Then, the element 8/+ is processed and node 5 and 8 are scheduled. Also, node 3 is discovered. Our new root list is {4/+,4/-,1/-,2/-,7/+,8/+,9/+,3/-}.
8. Then, element 9/+ is processed and node 9 is scheduled. New root list is {4/+,4/-,1/-,2/-,7/+,8/+,9/+,3/-}.

9. Finally, node 3 is processed and the algorithm is terminated.

We implemented the scheduling algorithm in C. The resulting schedule is given in Fig. 15.

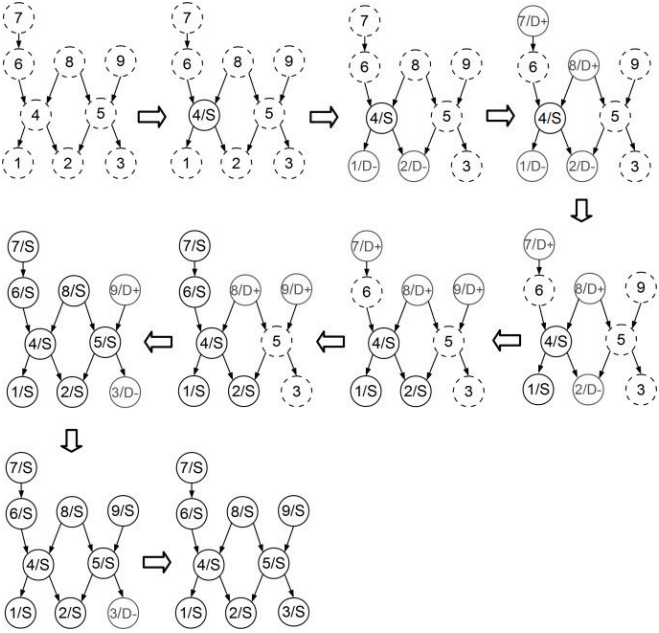

Fig. 14. An example that shows how the scheduling algorithm works.

```
1r   0         3f  -264      6f  -528
1f  -1828      4r  -1978     7r  -1642
2r   114       4f  -2770     7f  -3368
2f  -2092      5r  -1756
3r  -1564      6r  -1450
```

Fig. 15. Resulting timing schedule.

### D. Verification of the Schedule File

In this step, the timing schedule obtained is verified. It is compared against the constraint file for verification. This process is automatically carried out by a Perl script.

### E. Generating Timing Schedule SPICE Files

As it is mentioned before, the timing schedule specifies target values for each clock phase that will be given to the CTS tool. Before going through synthesis, it is beneficial to check these target values on the analog circuit. SPICE files for fast, slow, and typical PVT corners are automatically generated for simulation via a Perl script.

## V. DESIGN AND SYNTHESIS

In this section, we will give the details of our clock tree solution and steps that need to be followed to synthesize it. The flow of the synthesis process is given in Figure 16. We will first present the architecture of our clock tree solution, and then, will explain each synthesis step in a separate subsection.

### A. Proposed Clock Tree Architecture

The synthesized clock tree should generate the required clock phases and distribute these phases to the switches without violating the timing relationships between each other. Thus, we have a 2-tier clock tree solution that consists of two kinds of clock trees: *Clock Tree intrinsic* (*CTi*) and *Clock Tree extrinsic* (*CTe*). CTi generates the required clock phases from the main clock (clk) in regards to the timing constraints and schedule file. CTe distributes the generated clock phases to the clock pins of the analog circuit with preserving the timing relationships between the phases. In addition, CTe provides the inverse of a generated clock phase if it is required.

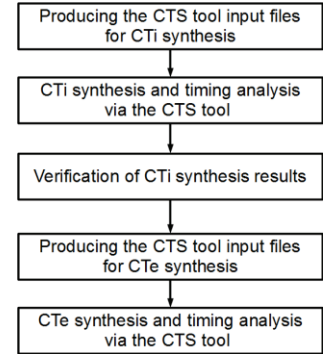

Fig. 16. Steps of Design and Synthesis.

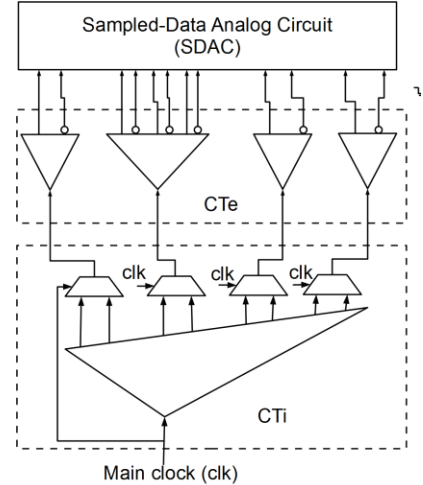

Fig. 17. Architecture of the proposed clock tree solution.

In Fig. 17, a diagram that depicts the roles of CTi and CTe can be observed. The triangles in the figure represent clock trees while the small circles represent inverters. Clock trees presented in the figure consist of buffers and inverters of our standard-cell design library.

Detailed info on CTi and CTe architectures will be given in the following subsections.

### B. Clock Tree intrinsic (CTi)

CTi uses a 2-to-1 multiplexer to generate each required clock phase as it is shown in Fig. 18. Each output clock phase ($clk_{out}$) is generated by combining two input phases ($clk_A$ and $clk_B$), which are obtained by simply delaying the main clock (clk), via the multiplexer. For instance, to generate a clock signal with period $T_{clk}$ and duty cycle ($T_{clk}/2 + \Delta t_A - \Delta t_B$), the phases $clk_A$ and $clk_B$ are generated first. The phase difference between these phases and the main clock are $\Delta t_A$ and $\Delta t_B$, respectively. Then, $clk_A$ is selected if clk is low and $clk_B$ is selected if clk is high as the output phase ($clk_{out}$).

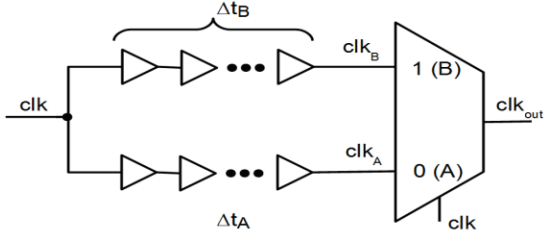

Fig. 18. Producing a clock phase from the main clock.

As it is known, each generated clock drives a switch in the analog circuit. And each switch consists of an NMOS and/or a PMOS transistor. Hence, each clock phase has two variants, which are the inverse of one another: the first variant (i.e., regular phase) drives NMOS transistors while the second one (i.e., complementary phase) drives the PMOS transistors. We name these two variants with "_clk" and "_xclk" suffixes, respectively.

An important point is that we should use the same edge of the main clock as the reference while synthesizing the clock edges of the constraint file. This requires generating regular clocks for some phases and the complementary clocks for the others. Considering the constraint (1r -> 3f), the edge 3f uses the falling edge of the main clock as the reference edge while 1r uses the rising edge of the main clock as the reference edge. However, generating regular clocks for both phases would cause wrong results as rising and falling delays of the multiplexer's ports are different. Instead, we should generate regular clock for one of the phases and the complementary clock for the other. In this paper, we choose the rising edge of the main clock as the reference edge. Therefore, we generate regular clock for phase 1, while we generate the complementary clock for phase 3.

Another important issue that should be addressed here is the exceptional case of generating clock phases, of which period values are $2 \times T_{clk}$. An SDAC generally requires phases with a duty cycle of $T_{clk}/2$ and a period of $T_{clk}$. However, in some cases, phases with a period of $2 \times T_{clk}$ may also be required. Even duty cycles might differ for some cases. In our test circuit, we have to use clocks, of which periods are $2 \times T_{clk}$, and duty cycles are $T_{clk}$ and $T_{clk}/2$. In order to generate these exceptional phases, we extended CTi by adding extra logic as it is shown in Fig. 19.

To be able to generate the clocks with $2 \times T_{clk}$ period, we should first generate a reference clock with $2 \times T_{clk}$ period. We use a one-bit counter, shown in Fig. 19a, to generate this reference clock. In Fig. 19b-d, we show the CTi structures to generate different clock phases:

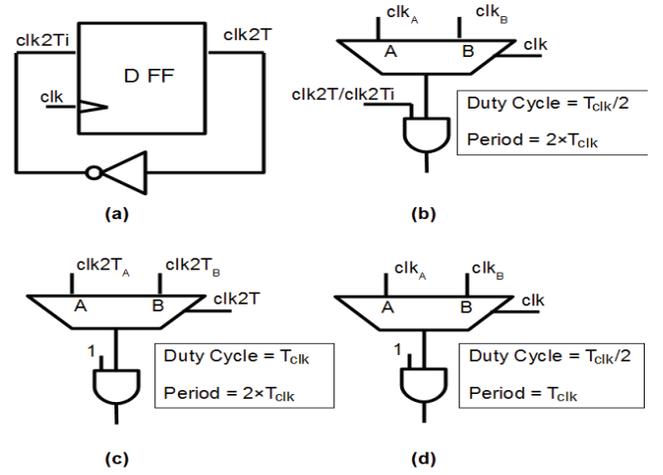

Fig. 19. CTi modification: (a) Generating the reference clock with $2 \times T_{clk}$ period. (b) Generating a ($T_{clk}/2$, $2 \times T_{clk}$)-clock. (c) Generating a ($T_{clk}$, $2 \times T_{clk}$)-clock. (d) Generating a ($T_{clk}/2$, $T_{clk}$)-clock.

- Fig. 19b shows how to generate a clock signal with a duty cycle of $T_{clk}/2$ and a period of $2 \times T_{clk}$.
- Fig. 19c shows how to generate a clock signal with a duty cycle of $T_{clk}$ and a period of $2 \times T_{clk}$.
- Fig. 19d shows how to generate a clock signal with a duty cycle of $T_{clk}/2$ and a period of $T_{clk}$.

The AND gates of Fig. 18c and d are functionally redundant. However, they are required to apply the delay overhead of the AND gate in Fig. 19b to each generated clock phase.

To sum up, we generate the rising and falling edges of each clock phase separately as delayed variants of the main clock. After generating the edges, we select the required edge using a multiplexer as shown in Fig. 18. In this way, we convert the clock phase generation problem of the SDACs into a form that can be understood by a digital CTS tool.

### C. Clock Tree extrinsic (CTe)

The role of CTe is to deliver the clock phases generated by CTi to the analog circuit without violating the timing relationships between phases. SDAC is treated as a 'hard macro' by CTe and it is not allowed to put a cell or metal wire within SDAC boundaries. Hence, CTe delivers the clock signals to just outside the SDAC boundaries, ensuring metals on both sides connect to each other. Each SDAC clock pin has a capacitance value of its own. Moreover, the delay between each clock pin and the switch associated with the pin is different. Since the clock trees inside CTe take these effects into account, they can be thought of as impedance transformers. In Fig. 20, we show a clock tree of CTe.

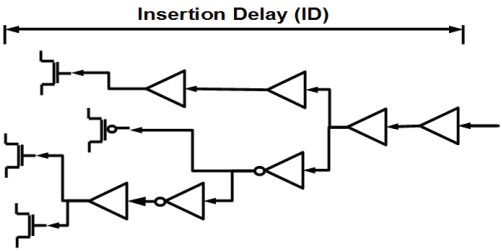

Fig. 20. A clock tree of CTe.

In CTe, we generate a clock tree for each CTi output, and every clock tree distributes its clock and the clock's complement to the switches on analog macro as in Fig. 20. Each clock tree in CTe is very much like the ones in synchronous digital systems, that is each clock tree tries to distribute the CTi generated clock signal to the related switches with the smallest skew possible.

### D. Generating Input Files for CTi Synthesis and Analysis

In order to drive the CTS tool for CTi synthesis we need the following files:
- A *command file*, which includes Tcl commands to drive the CTS tool,
- a *configuration file*, which acts as an interface between the CTS tool and the host computer,
- a *Verilog file*, which describes the structure of CTi circuit,
- *LEF* and *LIB files* of the standard cell, and
- a *clock tree specification file*. This file is specific to the CTS tool. It converts our *schedule file* into the format of our CTS tool. This conversion is done via a Perl script.

The LEF and LIB files of the standard-cell library are provided by the manufacturer. The other files are generated automatically by our Perl scripts.

### E. CTi Synthesis and Timing Analysis

In this step, CTi synthesis and analysis is done with the CTS tool (i.e., Cadence Encounter). This process is fully automated via a Perl script. The Perl script
- executes Cadence Encounter,
- synthesizes CTi by driving Cadence Encounter,
- does the timing analysis of the synthesized CTi for three PVT corners, and
- saves the timing results into associated files.

### F. Verification of CTi Synthesis

After synthesis, the resulting CTi timings should be verified with respect to the timing values in the constraint file. The verification of is automatically handled by two Perl scripts.

The first script skims through CTi synthesis result files and grabs rising delay, rising slew rate, falling delay, and falling slew rate for each clock phase.

The second script compares these values with the timing values of the constraint file, and writes the comparison results into associated files.

With the verification of CTi synthesis results, CTi design phase is complete and CTe synthesis can start.

### G. Generating Necessary Files for CTe synthesis

There are numerous files required for CTe synthesis similar to its CTi counterpart. Most of the files are generated automatically by our scripts. However, some of the analog information should be obtained from the analog designer. Luckily, most analog design software can generate these files automatically as well. The required files for CTe synthesis are listed below:
- A *pin location file*, which contains the locations of the analog circuit's clock pins. This file is generated by analog design tool.
- A *pin capacitance file*, which contains the capacitance values of the analog circuit's clock pins. This file is also generated by analog design tool.
- A *pin parasitic delay file*, which contains the parasitic delay values of the analog circuit's clock pins. This file is also generated by analog design tool.
- LEF file of CTi, which contains the pin geometries and locations for each CTi clock pin. It is automatically generated by a Perl script.
- LEF file of the analog circuit. This file contains layout information of the analog circuit's pins such as layer, location, and geometry information. It is automatically generated by a Perl script using the pin location file.
- LIB file of the analog circuit. It is automatically generated by a Perl script using the pin capacitance file.
- A *command file* to drive the CTS tool for CTe synthesis. It is generated by a Perl script.
- A *configuration file*, which is similar to its CTi counterpart.
- A *Verilog file*, which describes the structure of CTe circuit,
- A *clock tree specification file*, which describes CTe in a format that is specific to Cadence Encounter.

### H. CTe Synthesis and Analysis

Similar to its CTi counterpart, this process is fully automated and carried out by a Perl script.

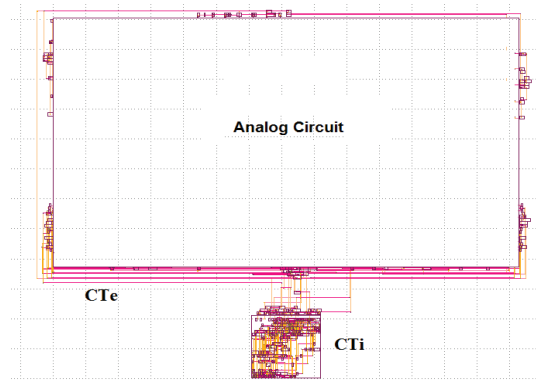

Fig. 21. Layout for CTi and CTe.

It is imported to note that CTi results are also taken into account during CTe synthesis. CTi clock pin locations as well as CTi clock rise and fall skew values are all supplied to Cadence Encounter for analysis. In this way, CTi circuit can be emulated during CTe synthesis.

Resulting layout after completion of CTe synthesis can be observed in Fig. 21.

## VI. Verification

Up to this point, CTi and CTe syntheses have been done. And our clock tree layout is ready for merging with the layout of the analog circuit. However, the synthesized clock tree should be verified by checking whether the design is compatible with the constraints or not before merging it with the SDAC. As it can be seen from Fig. 22, verification process consists of two steps.

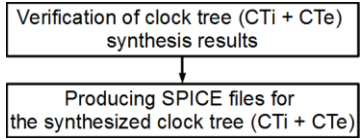

Fig. 22. Verification steps.

### A. Verification of Clock Tree Synthesis Results

Verification is done in three steps explained below.
- The resulting delay values from both CTi and CTe synthesis are added up and rising time, rising skew, falling time and falling skew values for each clock pin is obtained.
- Minimum and maximum edge values for each clock edge are calculated.
- Verification of the constraint file with the values obtained at step 2.

A clock phase, which is generated by CTi, may feed multiple clock pins of the analog circuit. Hence, there may be multiple CTe trees for a CTi output phase. In other words, a clock phase in the constraint file may correspond to multiple clock phases after synthesis. For instance, the phase $S_1$ (i.e., phase 1) of our CTi is connected to three clock pins of the analog circuit, namely, $S_1\_L_1\_clk$, $S_1\_L_1\_xclk$, and $S_1\_L_2\_xclk$. Similarly, the phase H (i.e., phase 3) is also connected to four clock pins of the analog circuit, namely, $H\_L_1\_clk$, $H\_L_2\_clk$, $H\_L_1\_xclk$, and $H\_L_2\_xclk$. In Fig. 23a, we list the obtained synthesis results for these variants in the following format:

<rising time> <rising slew> <falling time> <falling slew>

Here, rising time corresponds to ΔtB of Fig. 18 while falling time corresponds to ΔtA of Fig. 18. As it can be seen, different variants of the same CTi phase can have different rising and timing time values. In verification step, we should take this fact into account. For this purpose, we define a composite phase for each clock phase of the constraint file. A composite phase includes the effects of all variants of the corresponding CTi phase and has for edges: *minimum rising edge* ($r_{min}$), *maximum rising edge* ($r_{max}$), *minimum falling edge* ($f_{min}$), and *maximum falling edge* ($f_{max}$). Maximum/minimum rising edge is the maximum/minimum of the rising times of the variants. Similarly, maximum/minimum falling edge is the maximum/minimum of the falling times of the variants. All of the obtained composite phases are give in Fig. 23b in the following format:

<$r_{max}$> <$r_{min}$> <$f_{max}$> <$f_{min}$>

While computing if a constraint is satisfied, we should consider the worst case. In other words, we should select the most challenging composite phase edges for the constraint. As an example, one should choose '1$f_{max}$' and '3$r_{min}$' for the constraint '1f 3r 150'. Then, the constraint will become '1$f_{max}$ 3$r_{min}$ 150'. Using the values in Fig. 23b, the verification result will be '3120−2897=223', which satisfies the constraint. Since the constraint is checked for the worst case combination, it is evident that each variant will pass the verification. The verification results of all constraints of our test circuit are listed in Fig. 23c.

```
H_L1_clk  3128 38 4474 23    1 S1   4685 4668 2897 2865    1r 2r 0      Pass (101)
H_L1_xclk 4470 62 3122 54    2 S2   4818 4786 2587 2557    2f 1f 150    Pass (278)
H_L2_clk  3129 35 4475 22    3 H    3129 3120 4475 4468    1f 3r 150    Pass (223)
H_L2_xclk 4468 52 3120 44    4 CRes 2745 2739 1966 1960    3f 1r 150    Pass (193)
S1_L1_clk  4668 73 2865 61   5 Flop 2980 2963 3028 2982    4f 5r 900    Pass (997)
S1_L1_xclk 2896 58 4685 42   6 DlyIn 3285 3267 4187 4179   3r 6r 0      Pass (138)
S1_L2_xclk 2897 77 4685 52   7 DlyOut 3096 3026 1425 1378  6f 3f 150    Pass (281)
.                                                          5r 3f 150    Pass (1488)
.                                                          7r 6f 1000   Pass (1083)
                                                           5r 7r 0      Pass (46)
                                                           4r 3r 300    Pass (375)
                                                           6r 6f -1000  Pass (894)
                                                           2r 2f -5800  Pass (-2261)
                                                           7r 7f -4600  Pass (-1718)
                                                           3r 4f -3300  Pass (-1169)
       (a)              (b)                (c)
```

Fig. 23. Layout for CTi and CTe.

If each constraint passes the verification phase, the clock tree is generated successfully and it can be merged with the SDAC for analog verification.

### B. Verification of Clock Tree Synthesis Results

In this step, SPICE files for the clock tree (i.e., CTi + CTe) are generated for the three PVT corners. These files enable analog simulation of the analog circuit and the synthesized clock tree together. SPICE files are automatically generated by a Perl script. The script uses output files of the synthesis step as input.

## VII. Results and Conclusions

In this paper, we propose a methodology for automated design and synthesis of clock trees for SDACs. In our methodology, the analog designer only needs to connect switches to the clock pins of the SDAC and provide a few input files for our tools:
- A definition file that includes the duty cycle and period information for the required phases.
- A constraint file that specifies the timing constraints.
- Three files that include the clock pin locations,

capacitances, and parasitic delays of the SDAC.

After feeding the above files into our tools, the analog designer automatically obtains a clock tree for the SDAC. As a result, the design of the clock tree will be faster and less error-prone than the traditional approach, i.e., manual design.

In order to evaluate the performance of our clock tree synthesis methodology, we used the evaluation scheme given in Fig. 23. As it can be seen, the starting point of the evaluation scheme is the above input files, while the output is ENOB (Effective Number of Bits) value of our ADC circuit. We also compute SNR (Signal to Noise Ratio), SFRD (Spurious-Free Dynamic Range), and Energy per Bit.

ACTreS generates SPICE files for the targeted (Section IV) and synthesized (Section V) clock phases. In addition, a SPICE file for the test circuit is generated by the analog design tool. Then, simulation of the targeted and synthesized clock phases are done via an analog simulator. Finally, the performance values (i.e., ENOB values) are obtained by applying FFT analysis in MATLAB.

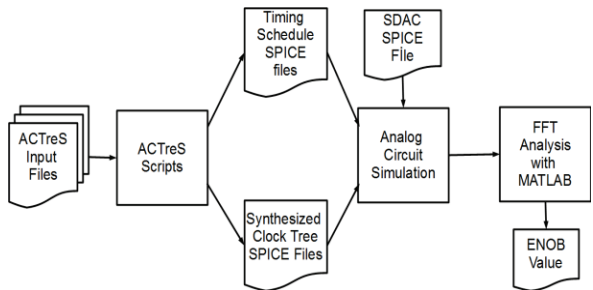

Fig. 24. Experimental setup.

We obtained the results listed in Table II using the setup of Fig. 24. As it can be seen from Table II, there is a 4% performance difference, on the average, between the synthesized and the targeted clock phases. Hence, we can state that we are able to synthesize a targeted clock tree for an SDAC with a slight performance difference using our methodology.

Note that if manual design is done, everything in this flow could be used except for the design and synthesis stage. The designer could still take our target clock schedule as a design goal and then use our verification step to check if his/her final design satisfies the constraints he initially set forward.

TABLE II
EVALUATION RESULTS FOR THE PROPOSED METHODOLOGY

| Type of the Phase | PVT corner | ENOB | SNR (dB) | SFDR (dB) | Energy per Bit (pJ/Cs) |
|---|---|---|---|---|---|
| Targeted | Slow | 7.63 | 47.69 | 57.35 | 6.31 |
| Synthesized | Slow | 7.00 | 43.90 | 56.14 | 9.76 |
| Targeted | Typical | 7.78 | 48.60 | 58.84 | 5.6 |
| Synthesized | Typical | 7.63 | 47.69 | 58.60 | 6.31 |
| Targeted | Fast | 6.61 | 41.55 | 55.64 | 12.79 |
| Synthesized | Fast | 6.75 | 42.27 | 55.16 | 11.77 |